\documentclass[12pt]{article}  
\usepackage{epsfig}
\def\3{\ss}                                                                     
\newcommand{\simge}{\raisebox{-1mm}{$\stackrel{>}{\sim}$}}                      

\begin{document}
\begin{center}
\begin{Large}
\begin{bf}                                                                 
  The Parameterized Simulation of \\
  Electromagnetic Showers in \\
  Homogeneous and Sampling Calorimeters \\
\end{bf}  
\end{Large}
\vspace*{10mm}
\begin{large}
  G.~Grindhammer\footnote{guenterg@desy.de} and S.~Peters \\
\end{large} 
\vspace*{5mm}
  Max-Planck-Institut f\"ur Physik \\
  (Werner-Heisenberg-Institut) \\
  F\"ohringer Ring 6 \\
  D-80805 M\"unchen 40, Germany 
\end{center}
\vspace*{5mm}
\begin{quotation}
\noindent
{\bf Abstract:}                                                                   
 A general approach to a fast simulation of electromagnetic showers using 
 parameterizations of the longitudinal and radial profiles in homogeneous 
 and sampling calorimeters is described. The dependence of the shower 
 development on the materials used and the sampling geometry is taken into 
 account explicitly. Comparisons with detailed simulations of various 
 calorimeters and with data from the liquid argon calorimeter of the H1 
 experiment are made.
\end{quotation}
\vspace*{5mm}
                                                                               
\section{Introduction}
                                                          
 In calorimeter simulation different tasks can be                               
 distinguished: calorimeter studies, physics analysis, and                       
 feasibility studies.                                             
 A detailed simulation, where all secondary particles are                        
 tracked individually down to some minimum energy and where the
 response is predicted from ``first principles'', is                            
 required for accurate calorimeter studies.                                    
 For physics analysis and feasibility studies large                             
 number of Monte Carlo events may have to be produced.                              
 Using individual particle tracking,                                            
 the computing time needed for such kind of simulations                         
 increases approximately linear with the energy absorbed in the                 
 detector and can easily become prohibitive.                                    
 Using parameterizations for electromagnetic (sub)showers                       
 can speed up the simulations considerably, without                             
 sacrificing precision.                                                         
 The high particle multiplicity in electromagnetic showers                      
 as well as their compactness and the good understanding of the                 
 underlying physics makes their parameterization advantageous.                  
                                                                                
 Using an Ansatz by Longo and Sestili \cite{longo},                             
 a simple algorithm for the description of longitudinal                         
 shower profiles has been used successfully                                      
 for the simulation of the UA1 calorimeter                                      
 \cite{bock}. Later, this Ansatz has been extended to the simulation            
 of individual showers, taking their shower-to-shower                           
 fluctuation and correlations                                                   
 consistently into account \cite{hayashide,badier,nim90}.                       
 For the parameterized simulation of radial energy profiles                     
 no conclusive procedure has been established until now.                        
                                                                                
 In homogeneous media, a scaling of the longitudinal                             
 and radial profiles in radiation lengths and Moli\`{e}re radii                
 respectively does not lead to a material independent description
 of electromagnetic shower development.                                            
 In sampling calorimeters, the shower shapes depend in addition                 
 on the sampling structure.                                                     
 We have extended the above Ansatz for parameterized simulation                 
 of longitudinal profiles by taking the material and geometry                   
 dependence of the parameters into account and developed a new algorithm        
 to simulate radial energy distributions \cite{diss}.                           
 Correlations between the longitudinal and radial shower                            
 development have been included.                                               
                                                                               
\section{Procedure}
                                                             
 To arrive at a general description of electromagnetic shower                   
 development, we performed detailed Monte Carlo                                   
 simulations, on a grid of 1.0 $X_0$ in depth and 0.2 Moli\`{e}re radii
 laterally, for various homogeneous media and sampling                           
 calorimeters, using the GEANT package \cite{geant}. The materials used
 were Cu, Fe, W, Pb, U, and scintillator and liquid argon.                   
 In a first step only average shower profiles in homogeneous                     
 media were analyzed, from which scaling laws for the material and         
 energy dependence of the parameters have been extracted.                       
 Starting from the relations which describe the average                         
 behavior of the parameterized quantities, we developed                        
 parameterizations for individual electromagnetic showers in homogeneous         
 calorimeters, taking fluctuations and correlations into                        
 account.                                                                       
                                                                                
 The parameterizations in homogeneous media are a first                          
 approximation for electromagnetic shower development in                        
 sampling calorimeters, which are viewed as consisting of                     
 one single effective medium.
 The inhomogeneous material distribution in sampling calorimeters
 influences however the exact behavior of the shower
 shapes. This is explained mainly by the transition effect which
 depends on the shower depth \cite{diss,flauger,wigmans}.
 These effects have been taken into account by adding geometry dependent
 terms to the parameterizations for homogeneous media, which
 can be easily calculated from the sampling geometry.                          
                                                                               
\section{Parameterization Ansatz}
                                                  
 The spatial energy distribution of electromagnetic showers is given by                            
 three probability density functions (pdf),                                     
 \begin{equation}                                                               
  dE (\vec{r}) \, = \, E\, f(t)dt\, f(r)dr\, f(\phi) d\phi ,                    
 \end{equation}                                                                 
 describing the longitudinal, radial, and azimuthal energy                       
 distributions.                                                                 
 Here $t$ denotes the longitudinal shower depth in units of                     
 radiation length, $r$ measures the radial distance from the shower            
 axis in Moli\`{e}re units, and $\phi$ is the azimuthal angle.                  
 The start of the shower is defined by the space point, where                   
 the first electron or positron bremsstrahlung process occurs.                 
 A gamma distribution is used for the parameterization                    
 of the longitudinal shower profile, f(t).                                      
 The radial distribution, f(r),                                                 
 is described by a two-component Ansatz. In $\phi$, it is assumed
 that the energy is distributed uniformly:                            
 $ f(\phi) = 1/2\pi $.

 \subsection{Longitudinal shower profiles -- homogeneous media}                  
 \label{sec_hom_long}
                                                           
 It is well known that average longitudinal shower profiles can                
 be described by a gamma dis\-tri\-bution \cite{longo}:                       
  \begin{equation}                                                              
    \left\langle \frac{1}{E} \frac{dE(t)}{dt} \right\rangle                     
                                 \, = \, f(t) \, = \,                           
    \frac{ (\beta t)^{\alpha -1} \beta \exp(-\beta t) }                         
         { \Gamma(\alpha) }.                                                    
  \end{equation}                                                                
 The center of gravity, $\langle t \rangle$, and the depth of the               
 maximum, $T$, can be calculated from the shape parameter $\alpha$              
 and the scaling parameter $\beta$ according to                                 
\begin{eqnarray}                                                                
  \langle t \rangle  & = & \frac{\alpha}{\beta}                   \\            
  T                  & = & \frac{\alpha-1}{\beta} .                             
  \label{talp}                                                                  
\end{eqnarray}                                                                  
 Longitudinal electromagnetic shower development in homogeneous                  
 media had been studied analytically by Rossi \cite{rossi}. 
 An important result of the calculations using                        
 ``Rossi Approximation B`` is that longitudinal shower 
 moments are equal in different materials, provided one 
 measures all lengths in units of radiation length $(X_0)$ 
 and energies in units of the critical energy          
 ($E_c$). Numerically, $E_c$ can be calculated according to \cite{dov}
 \begin{equation}                                                               
    E_c \, = \, 2.66 \left( X_0 \frac{Z}{A} \right)^{1.1} .                      
  \label{ec}                                                                    
 \end{equation}                                                                 
 For the depth of the shower maximum                                           
 \begin{eqnarray}                                                                
   T \, \propto \, \ln y \, = \, \ln \frac{E}{E_c}                              
 \end{eqnarray}                                                                  
 is predicted \cite{rossi}.                                                     
                                                                                
 It is therefore desirable to use $T$ in the parameterization. This               
 is demonstrated in Fig.\ref{p_thom}, where the average depth of            
 the shower maximum for various homogeneous media\footnote{                       
  The index ``hom`` in the following formulae indicates the                     
  validity for homogeneous media. For sampling calorimeters the                  
  index ``sam`` will be used.}, $T_{hom}$, is                                   
 plotted versus $y$, in the energy range from 1 to 100 GeV.                      
 As a second variable $\alpha$ is used. In this case the                     
 parameterization depends on the charge number $Z$ of the                     
 medium, as can be seen in Fig.\ref{p_ahom}. The lines in both figures
 correspond to fits to GEANT simulations according to                           
  \begin{eqnarray}                                                              
 \label{e_thom}                                                                 
   T_{hom} & = & \ln y + t_1 \\                                                 
 \label{e_ahom}                                                                 
   \alpha_{hom} & = & a_1 + (a_2 + a_3/Z) \ln y.                                
  \end{eqnarray}                                                                
 The values of the coefficients are given in Appendix,
 where all formulae and numbers, which will be given in the                  
 following, are summarized.                                                      
                                                                                
 Assuming that also individual profiles can be approximated by                  
 a gamma distribution, the fluctuations and correlations                        
 can be taken into account consistently                                         
 (for details refer to \cite{nim90}).                                      
 For each single GEANT-simulated shower, $T$ and $\alpha$ are determined          
 by fitting a gamma distribution. The logarithms                   
 of $T$ and $\alpha$ are used for the parameterization since they
 are found to be approximately normal distributed.                             
 For the parameterization of $\langle \ln T_{hom} \rangle$ and
 $\langle \ln \alpha_{hom} \rangle$                                             
 the logarithms of equations \ref{e_thom} and \ref{e_ahom}                      
 are used. The $y$-dependence of the fluctuations can be described
 by                                                                   
  \begin{equation}                                                              
   \sigma \, = \, ( s_1 + s_2 \ln y )^{-1} .                                     
   \label{lsighom}                                                              
  \end{equation}                                                                
 The correlation between $\ln T_{hom} $ and $\ln \alpha_{hom} $
 is given by                                                                    
  \begin{equation}                                                              
   \rho(\ln T_{hom}, \ln \alpha_{hom}) \, \equiv \, \rho                        
                                       \, = \, r_1 + r_2 \ln y .                
   \label{corrhom}                                                              
  \end{equation}                                                                
 The dependence of these quantities on $y$ is shown in                          
 Fig.\ref{fitlhom} for various materials together with                       
 the parameterizations (see Appendix \ref{a_hom_long_ind}).
                                                                                
 From these formulae, correlated and varying parameters                             
 $\alpha_i$ and $\beta_i$ are generated according to                          
\begin{equation}                                                                
  \left(                                                                        
     \begin{array}{c}                                                           
       \ln T_i \\                                                               
       \ln \alpha_i                                                             
     \end{array} \right)                                                        
    \, = \,                                                                     
  \left(                                                                        
     \begin{array}{c}                                                           
       \langle \ln T \rangle \\                                                 
       \langle \ln \alpha \rangle                                               
     \end{array} \right)                                                        
    + C                                                                         
  \left(                                                                        
     \begin{array}{c}                                                           
       z_1 \\                                                                   
       z_2                                                                      
     \end{array} \right)                                                        
\end{equation}                                                                  
 with                                                                           
$$                                                                              
  C \, = \,                                                                     
     \left(                                                                     
     \begin{array}{cc}                                                          
      \sigma (\ln T) & 0 \\                                                     
      0 & \sigma (\ln \alpha)                                                   
     \end{array} \right)                                                        
     \left(                                                                     
     \begin{array}{cc}                                                          
       \sqrt{\frac{1+\rho}{2}} & \sqrt{\frac{1-\rho}{2}} \\                     
       \sqrt{\frac{1+\rho}{2}} & - \sqrt{\frac{1-\rho}{2}}                      
     \end{array} \right) \,                                                     
$$                                                                              
 and $\beta_i = (\alpha_i - 1)/T_i$ and $z_1$ and $z_2$ are standard             
 normal distributed random numbers.                                             
 The longitudinal energy distribution is evaluated\footnote{~The GAMDIS function 
 of the CERN computer library is used.}
 by integration              
 in steps of $\Delta t = t_j - t_{j-1} = 1 X_0$,                                
 $$                                                                             
   dE(t) = E \int_{t_{j-1}}^{t_j}                                               
         \frac{ (\beta_i t)^{\alpha_i -1} \beta_i \exp(-\beta_i t) }            
              { \Gamma(\alpha_i) } dt \, .                                      
 $$
 It is worthwhile to mention that only one                              
 of the five quantities needed, $\langle \ln \alpha_{hom} \rangle$,
 depends explicitly on the material, while for the other four this dependence 
 is absorbed by using $y$ instead of $E$.                                                   
                                                                                
 In Fig.\ref{longsf5} longitudinal profiles of GEANT and                     
 parameterized simulations for a lead glass calorimeter (SF5)                      
 are compared. Shown are the mean profiles and the mean + 1 RMS                 
 in each $X_0$ interval. While the means are in perfect agreement,                
 the fluctuations are underestimated by the parameterized                       
 simulations at low energies, indicating that the description of                
 individual profiles by gamma distributions becomes a worse                    
 approximation with decreasing shower energy.                                   
 Comparisons for other materials (Fe, Cu, W, Pb, U) are of                      
 comparable quality as those in Fig.\ref{longsf5} \cite{diss}.                     
 In the next sections we will show, how the sampling fluctuations in sampling                        
 calorimeters can be used to improve the shape fluctuations at low energies.

 \subsection{Sampling fluctuations}                                             
 \label{sec_samp_fluc}
                                                          
 In fast simulations, sampling calorimeters consisting of a                     
 complicated but repetitive sampling structure are usually described            
 by one single effective medium (the formulae to compute effective
 material parameters are summarized in Appendix \ref{a_sam_mat}).               
 The sampling fluctuations, the scaling of the deposited energy 
 to the visible energy using an appropriate sampling fraction, and 
 the effects of the sampling structure have to be considered
 in parameterized simulations explicitly.

 The simulation of sampling fluctuations are                                        
 done conveniently with a gamma distribution:
 \begin{equation}                                                               
   G(a,b) \, = \, \frac{x^{a-1}be^{-bx}}{\Gamma (a)}                            
  \label{gam_fluc}                                                              
 \end{equation}                                                                 
 with                                                                           
\begin{equation}                                                                
  \langle x \rangle \, = \, \frac{a}{b} \, \, , \, \,                           
    \sigma^2(x) \, = \, \frac{a}{b^2} .                                         
\end{equation}                                                                  
 The energy in each longitudinal integration step, $dE(t)$,                    
 is fluctuated\footnote{~The RANGAM function of the CERN computer 
 library is used.}
 according to equation \ref{gam_fluc} choosing                    
 \begin{equation}                                                               
  a\, = \, \frac{dE(t)}{c^2} \mbox{ and } b\, = \, \frac{1}{c^2} .                
 \end{equation}                                                                 
 It is then easy to show that the central limit theorem will                       
 ensure the total energy to be normal distributed obeying the                   
 usual formula for the sampling fluctuations:
\begin{equation}                                                                
  \frac{\sigma}{E} \, = \, \frac{c}{\sqrt{E}}.                                  
\end{equation}                                                                  
 Using this procedure the occurrence of negative energies is                     
 automatically avoided. Additional fluctuations of the longitudinal
 shape are introduced, leading to a better agreement                       
 in the shape fluctuations.                                                     
 This method is also used to fluctuate energy                       
 depositions of real particles (electrons, hadrons), when they                      
 are tracked individually through an effective homogeneous                       
 volume.

 \subsection{Longitudinal shower profiles -- sampling calorimeters}             
 \label{sec_samp_long}
                                                          
 The inhomogeneous material distribution in sampling                             
 calorimeters influences                                                        
 the exact behavior of the shower shapes.                                      
 In the first stages of electromagnetic shower development the                  
 signal is dominated by electrons and positrons. Behind the                     
 shower maximum low energetic photons become more and more                      
 important. The transition effect, being explained mainly by                   
 the absorption properties of low energetic photons, must in                    
 turn depend on the shower depth. Consequently, the signal ratio                
 of electrons to minimum ionizing particles, $e/mip$, decreases continuously 
 as the shower propagates longitudinally. Thus the signal maximum                  
 in a sampling calorimeter occurs at an earlier depth                     
 than expected for a homogeneous calorimeter with the same                      
 effective material properties. This can be seen                                
 from Fig.\ref{fitlsam}                                                      
 (left upper corner), where $\langle \ln T \rangle$                             
 for homogeneous media is compared to the values in                              
 five different sampling calorimeters. In addition, the amount of the                
 shift of $\langle \ln T \rangle$ depends on the exact geometrical arrangement.             
                                                                                
 The parameterization of the longitudinal shape as given in                    
 section \ref{sec_hom_long} for homogeneous media can therefore                    
 not be used for sampling calorimeters directly. Instead it may                 
 be understood as a first approximation to which geometry                       
 dependent corrections have to be added.                                        
 We use the sampling frequency                                            
 \begin{equation}                                                               
   F_S \, = \, \frac{X_{0,eff}}{d_a+d_p}                                        
 \end{equation}                                                                 
 and the value of $e/mip$ (averaged over the shower depth) to
 account for the shower depth dependence of the transition effect.              
 $d_a$ and $d_p$ denote the thickness of the active and passive                 
 layers, respectively. If $e/mip$ is not known, a sufficiently good
 approximation for many calorimeters \cite{delpeso}          
 with charge numbers $Z_p$ and $Z_a$ is given by              
 \begin{equation}                                                               
   \hat{e} \, = \, \frac{1}{1+0.007(Z_p-Z_a)}\, \approx \, \frac{e}{mip} .        
  \label{ehatpes}                                                               
 \end{equation}                                                                 
 Averaged over the whole shower, $e/mip$ remains energy                        
 independent for $E \simge 1$ GeV.
             
 The average longitudinal profiles can now be parameterized according to
\begin{eqnarray}                                                                
   \label{tsam}                                                                 
   T_{sam} & = & T_{hom} + t_1 F_S^{-1} + t_2 (1-\hat{e}) \\                    
   \label{asam}                                                                 
   \alpha_{sam} & = & \alpha_{hom} + a_1 F_S^{-1} ,                               
\end{eqnarray}                                                                  
 and the quantities used for the simulation of individual showers are given by
\begin{eqnarray}                                                                
   \label{exptsam}                                                              
   \langle \ln T_{sam} \rangle & = &                                            
      \ln \left(  \exp(\langle \ln T_{hom} \rangle )                            
                                          + t_1 F_S^{-1} + t_2 (                
               1 - \hat{e}) \right) \\                                          
   \label{expasam}                                                              
   \langle \ln \alpha_{sam} \rangle & = &                                       
      \ln \left( \exp( \langle \ln \alpha_{hom} \rangle )                       
                   + a_1 F_S^{-1} \right).                                      
\end{eqnarray}                                                                  
 The fluctuations, $\sigma(\ln T_{sam})$,
 $\sigma(\ln \alpha_{sam})$ and the correlation,
 $\rho (\ln T_{sam},\ln \alpha_{sam} )$,                                         
 are described with the help of the same formulae as in the case of             
 homogeneous media (see Appendix \ref{a_sam_long_ave} and                       
 \ref{a_sam_long_ind}).                                                         
                                                                                
 Fig.\ref{fitlsam} summarizes the parameterization for                      
 sampling calorimeters. The expectation value of $\ln T$ no                     
 longer scales with $y$. The expectation value of $\ln \alpha$
 depends on the material and the sampling geometry.                          
 The fluctuations and correlations of the parameters can still                  
 be approximated without any explicit material or geometry                      
 dependence.                                                                    
                                                                                
 In Figs.\ref{clinif} to \ref{cllsif} GEANT                  
 and parameterized simulations of the lead liquid argon                      
 calorimeter (IFE) of the H1 experiment \cite{h1prop,diss} are compared. 
 The GEANT simulations were performed                   
 with low energy cuts ($e$-cut$=200$ keV, $\gamma$-cut$=10$ keV)
 and a detailed geometry description, including for example              
 copper pads and G10 layers. These simulations were {\sl not} used to tune 
 the parameterizations.
 Both, average longitudinal profiles and their fluctuations                     
 (including sampling fluctuations) are in very good agreement                   
 (see Fig.\ref{clinif}). The energy containment                                  
 (see Fig.\ref{cllcif}) and the energy resolution (see 
 Fig.\ref{cllsif}) as a function of the longitudinal calorimeter
 length are also well predicted. Comparisons with detailed simulations of 
 other calorimeters (Fe-LAr, Cu-Sc, W-LAr, Pb-LAr, U-Sc) show a comparably 
 good performance \cite{diss}.

 \subsection{Radial shower profiles -- homogeneous media}                        
 \label{sec_hom_rad}                                                            

 Average radial energy profiles,                                          
 \begin{equation}                                                               
   f(r) \, = \, \frac{1}{dE(t)} \frac{dE(t,r)}{dr},                             
 \end{equation}                                                                 
 at different shower depths in pure uranium are presented in 
 Fig.\ref{hradx1}.                                         
 These profiles show a distinct maximum in the core of the shower                 
 which vanishes with increasing shower depth. In the tail                       
 ($r \simge 1 R_M$) the distribution looks nearly flat                          
 at the beginning ($1-2 X_0$), becomes steeper at moderate depths               
 ($5-6 X_0$, $13-14 X_0$), and becomes flat again                               
 ($22-23 X_0$).                                                                 
 A variety of different functions can be found in the literature to
 describe radial profiles \cite{ako77,abs79,fer88,peso89,sicapo,nim90}. 
 We use the following two component Ansatz, an extension of \cite{nim90}:
\begin{eqnarray}                                                                
  \label{frad}                                                                  
   f(r) &  = &  p f_C(r) + (1-p) f_T(r) \\                                      
              &  = &                                                            
                         p \frac{2 r R_C^2}{(r^2 + R_C^2)^2}                    
                    + (1-p) \frac{2 r R_T^2}{(r^2 + R_T^2)^2}                   
  \nonumber                                                                     
\end{eqnarray}                                                                  
 with                                                                           
  $$ 0 \leq p \leq 1 . $$                                                       
 Here $R_C$ ($R_T$) is the median of the core (tail) component                  
 and $p$ is a probability giving the relative weight of                        
 the core component. For the shower depth $1-2 X_0$ the distributions                              
 $f(r)$, $p f_C(r)$, and $(1-p) f_T(r)$ are also indicated in Fig.\ref{hradx1}.

 The evolution of $R_C$, $R_T$, and $p$ with increasing shower                   
 depth is shown in Fig.\ref{hrzrkp}                                          
 for 100 GeV showers in iron and uranium.                                       
 We use the variable $\tau = t/T$, which measures the                      
 shower depth in units of the depth of the shower maximum,                       
 to generalize the radial profiles. This makes the                               
 parameterization more convenient and separates the energy                      
 and material dependence of various parameters.                                 
 The median of the core distribution, $R_C$, increases linearly                 
 with $\tau$. The weight of the core, $p$, is maximal                         
 around the shower maximum, and the width of the tail,                           
 $R_T$, is minimal at $\tau \approx 1$. This behavior can                      
 be traced back to the radial profiles shown in Fig.\ref{hradx1}.
                                                                                
 The following formulae are used to parameterize the radial                   
 energy density distribution for a given energy and material:
\begin{eqnarray}                                                                
    \label{rz}                                                                  
 R_{C,hom}(\tau) & = &                                                          
    z_1 + z_2 \tau \\                                                           
    \label{rk}                                                                  
 R_{T,hom}(\tau) & = &                                                          
    k_1 \{ \exp (k_3(\tau -k_2)) + \exp (k_4(\tau -k_2)) \} \\                  
    \label{p}                                                                   
   p_{hom}(\tau) & = &                                                          
    p_1  \exp \left\{ \frac{p_2-\tau}{p_3} -                                    
              \exp \left(   \frac{p_2-\tau}{p_3} \right) \right\}               
\end{eqnarray}                                                                  
 The parameters $z_1 \cdots p_3$ are either constant or simple                  
 functions of $\ln E$ or $Z$ (see Appendix \ref{a_hom_rad_ave}                  
 for details). The complicated evolution of $R_T$ and $p$ with                  
 the shower depth and the dependence on the material can                        
 be explained mainly with the propagation of low energetic                      
 photons \cite{diss}. The offset in $R_T$ between iron and                      
 uranium (Fig.\ref{hrzrkp}) for example, indicating                          
 a wider distribution in iron, reflects the difference in the mean free path, 
 which for 1 MeV photons is approximately twice as long in iron as in uranium, 
 if lengths are measured in Moli\`{e}re units.
                                                                                
 We found a good agreement of mean radial profiles between                      
 parameterized and detailed simulations in Fe, Cu, W, Pb, and                    
 U absorbers for energies between 0.4 and 400 GeV. This is demonstrated in 
 Fig.\ref{crinhom3}, where radial profiles in various shower depths are                             
 compared for 40 GeV showers in lead and 100 GeV showers in uranium.
                                                                                
 The introduction of radial shape fluctuations has to be considered with some 
 care. Even if no fluctuations of $f(r)$ are simulated explicitly, the radial 
 energy profile at a given shower depth will fluctuate, because the shower                               
 maximum $T$ and thus $\tau$ varies from shower to shower. Another source of 
 radial fluctuations arises from the method, which we have adopted for the                                  
 simulation of radial distributions. The energy content of a longitudinal 
 interval of length 1 $X_0$, $dE(t)$, is calculated from the actual longitudinal 
 energy density distribution as described in section \ref{sec_hom_long}. This 
 energy is divided into $N_S(t)$ discrete spots of energy 
 $E_S\, =\, dE(t)/N_S(t)$, which are distributed radially according to $f(r)$ 
 using a Monte Carlo method. This can be done easily since the pdfs, $f_C(r)$ 
 and $f_T(r)$, can be integrated and inverted:
 \begin{eqnarray}                                                               
  F(r) & = & \int_{0}^{r} \frac{ 2 r^{\prime} R^{2} }                           
   { \left( r^{\prime 2} + R^{ 2} \right)^2} dr^{\prime}                        
   \, = \, \frac{r^2}{r^2 + R^2} \\                                             
  F^{-1}(u) & = & R \sqrt{ \frac{u}{1-u} } \, .                                 
 \end{eqnarray}                                                                 
 Random radii are generated according to $f(r)$ in the following way,
 using two normal distributed random numbers $v_i$ and $w_i$:
  $$ r_i    \, = \, \left\{                                                     
  \begin{array}{ll}                                                             
   R_C\sqrt{\frac{v_{i}}{1-v_{i}}}, & \mbox{if $p < w_{i}$} \\                  
   R_T\sqrt{\frac{v_{i}}{1-v_{i}}}, & \mbox{else.}                              
  \end{array} \right.  $$                                                       
                                                                                
 This method leads to additional fluctuations in the energy content of every 
 radial interval which follow a binomial distribution. Thus, the relation
 \begin{equation}                                                               
  \frac{\sigma^2(\epsilon)}{\langle \epsilon \rangle                            
    (1-\langle \epsilon \rangle )}                                              
  \, = \, const \, = \, N_S^{-1}                                                
 \label{g_sigbye}                                                               
 \end{equation}                                                                 
 describes the contribution to radial shape fluctuations                        
 produced by the Monte Carlo method in each longitudinal                        
 integration interval. Here $\epsilon$ denotes the energy in a given
 radial interval at a given shower depth:                                      
 \begin{equation}                                                               
  \langle \epsilon \rangle \, \equiv \,                                         
    \int_{r_1}^{r_2} f(r) dr \, = \,                                            
   \frac{dE(t,r)}{dE(t)} .                                                      
 \end{equation}                                                                 
                                                                                
 We investigated the possibility to tune $N_S(t)$ in                          
 each longitudinal interval to match the radial shape                           
 fluctuations observed in detailed GEANT 
 simulations\footnote{De Angelis et al.~have used                               
 a similar method to reproduce shape fluctuations                               
 \cite{deangelis}.}. As an example, the quantity              
 $ \sigma^2(\epsilon)/(\langle \epsilon \rangle                                 
    (1-\langle \epsilon \rangle )) $                                            
 at $t$ = $5-6 X_0$ is displayed in Fig.\ref{sigbyew} for 
 detailed simulations and parameterized ones without any radial shape                             
 fluctuations. The difference of these curves, which is also shown in 
 Fig.\ref{sigbyew}, is approximately constant and determines
 $N_S^{-1}$ in equation \ref{g_sigbye} (note that the                           
 variance is additive).                                                          
 We found that a constant contribution to                                      
 $ \sigma^2(\epsilon)/(\langle \epsilon \rangle                                 
    (1-\langle \epsilon \rangle )) $                                            
 can be used to match the total radial shape fluctuations to                    
 a good approximation at all shower depths.                                     
                                                                                
 Summing $N_S(t)$ over all shower depth,                                          
 the total number of spots, $N_{Spot}$, needed for one shower                     
 can be obtained and parameterized according to                                 
  \begin{equation}                                                              
     N_{Spot} \, = \, 93 \ln(Z) E^{0.876}.                                      
  \end{equation}                                                                
 To find the number of spots for each longitudinal integration              
 interval, the density distribution
 $1/N_{Spot}\, dN_S(t)/dt$ in Fig.\ref{sigbyew} is parameterized.               
 It is described by a gamma distribution with parameters,
 which are given by the corresponding longitudinal energy profile:
 \begin{eqnarray}                                                               
  T_{Spot} & = & T_{hom} ( 0.698 + 0.00212 Z ) \; \; \; \mbox{and} \\           
  \alpha_{Spot} & = & \alpha_{hom} ( 0.639 + 0.00334 Z ) .                      
 \end{eqnarray}                                                                 
 The total fluctuations obtained with this method are compared in 
 Fig.\ref{crinhom3} by adding 1 RMS to the mean profiles.                                             
                                                                                
 Additional correlations between longitudinal and radial                        
 shower development are taken into account by                                   
 introducing a correlation between the radial pdfs and                         
 the actual center of gravity,                                                  
$$ \langle t \rangle_i \, = \, \frac{\alpha_i}{\beta_i} \, = \,                 
   T_i \frac{\alpha_i}{\alpha_i - 1}, $$                                        
 of an individual shower. This is done by replacing $\tau$ in                 
 equations \ref{rz}, \ref{rk}, and \ref{p} by $\tau_i$:                                    
 \begin{equation}                                                               
  \tau \, = \, \frac{t}{T} \, \longrightarrow \,                                
   \tau_i \, = \, \frac{t}{\langle t \rangle_i}                                 
    \frac{\exp(\langle \ln \alpha \rangle )}                                    
         {\exp(\langle \ln \alpha \rangle )-1}\, .                              
 \end{equation}                                                                 
 The need to introduce these correlations is demonstrated in                    
 Fig.\ref{radsf5}, where integrated radial profiles are shown,               
 which were calculated by summing over all longitudinal layers.                   
 Note that the mean integrated profiles,                                        
 $$ \left\langle \frac{1}{E} \frac{dE(r)}{dr} \right\rangle \, ,$$              
 are independent of energy, which is well reproduced by the                          
 parameterized simulation. The relative fluctuations of these distributions,
 $$                                                                             
  \hat{\sigma}(r) \, \equiv \, \frac{ \sigma_{RMS} }                            
  { \left\langle \frac{1}{E} \frac{dE(r)}{dr} \right\rangle } \, ,                  
 $$                                                                             
 are shown using both, $\tau$ and $\tau_i$, in calculating the                    
 radial profiles. Only the simulations using $\tau_i$ are                       
 able to predict the fluctuations observed with GEANT correctly.
                                                                                
 For clarity, we summarize the steps of our algorithm as follows:
 Determine the energy $dE(t)$ within one longitudinal                           
 integration interval as described in section \ref{sec_hom_long}.               
 In case of sampling calorimeters apply sampling fluctuations                   
 on $dE(t)$. Evaluate the number of spots needed to reproduce radial shape 
 fluctuations in this interval according to
 $$                                                                             
   N_S(t) = N_{Spot} \int_{t_{j-1}}^{t_j}                                       
   \frac{ (\beta_{Spot} t)^{\alpha_{Spot} -1}                                   
            \beta_{Spot} \exp(-\beta_{Spot} t) }                                
        { \Gamma(\alpha_{Spot}) } dt \, .                                       
 $$                                                                             
 Distribute the spots with energy $E_S=dE(t)/N_S(t)$ radially according 
 to $f(r)$ as described above and uniformly in $\phi$ and in the
 longitudinal interval $\Delta t$. 
 Finally transform the spot coordinates $(E_S,t[X_0],r[R_M], \phi)$
 into the detector reference system $(E_S,x,y,z)$.
                                                                               
 \subsection{Radial shower profiles -- sampling calorimeters}                   
 \label{sec_samp_rad}                                                           
                                                                                
 The influence of the exact geometry on radial energy profiles                  
 is rather small. At the start of the shower the profiles look                  
 a bit smoother than in homogeneous media. With increasing shower                  
 depth they approach the shapes that are expected for                           
 homogeneous media with the appropriate effective material.                  
 These small deviations have been taken into account by the                     
 following corrections to the mean profiles:                                    
 \begin{eqnarray}                                                               
  R_{C,sam} & = & R_{C,hom} +                                                   
     z_1      ( 1-\hat{e} )                                                     
     + z_2 F_S^{-1} \exp( -\tau_i) \\                                           
  R_{T,sam} & = & R_{T,hom} +                                                   
     k_1 ( 1-\hat{e} )                                                          
     + k_2 F_S^{-1} \exp( -\tau_i) \\                                           
  p_{sam} & = & p_{hom} +                                                       
     ( 1-\hat{e} )                                                              
          ( p_1 + p_2 F_S^{-1} \exp(-(\tau_i-1)^2)  )                           
 \end{eqnarray}                                                                 
 using again the sampling frequency $F_S$ and $e/mip$ (see                 
 Appendix \ref{a_sam_rad_ave}).                                                 
                                                                                
 The total number of spots needed to simulate the radial                        
 shape fluctuations is much smaller than in the case of                         
 homogeneous media and no longer depends sensitively on                       
 the materials used. Instead, the spot number can be                            
 parameterized by                                                     
 \begin{equation}                                                               
    N_{Spot} \, = \, \frac{10.3}{c} E^{0.959} ,                                   
 \end{equation}                                                                 
 where $c$ measures the sampling fluctuations according to                      
  $$ \frac{\sigma}{E} \, = \, \frac{c}{\sqrt{E}}  $$                            
 (see Fig.\ref{nspots}).                                                     
 The density distribution of the spot numbers is given in               
 analogy to the homogeneous media by:                                               
\begin{eqnarray}                                                                
  T_{Spot} & = & T_{sam} ( 0.831 + 0.0019 Z ) \; \; \; \mbox{and} \\            
  \alpha_{Spot} & = & \alpha_{sam} ( 0.844 + 0.0026 Z )                         
\end{eqnarray}                                                                  
                                                                                
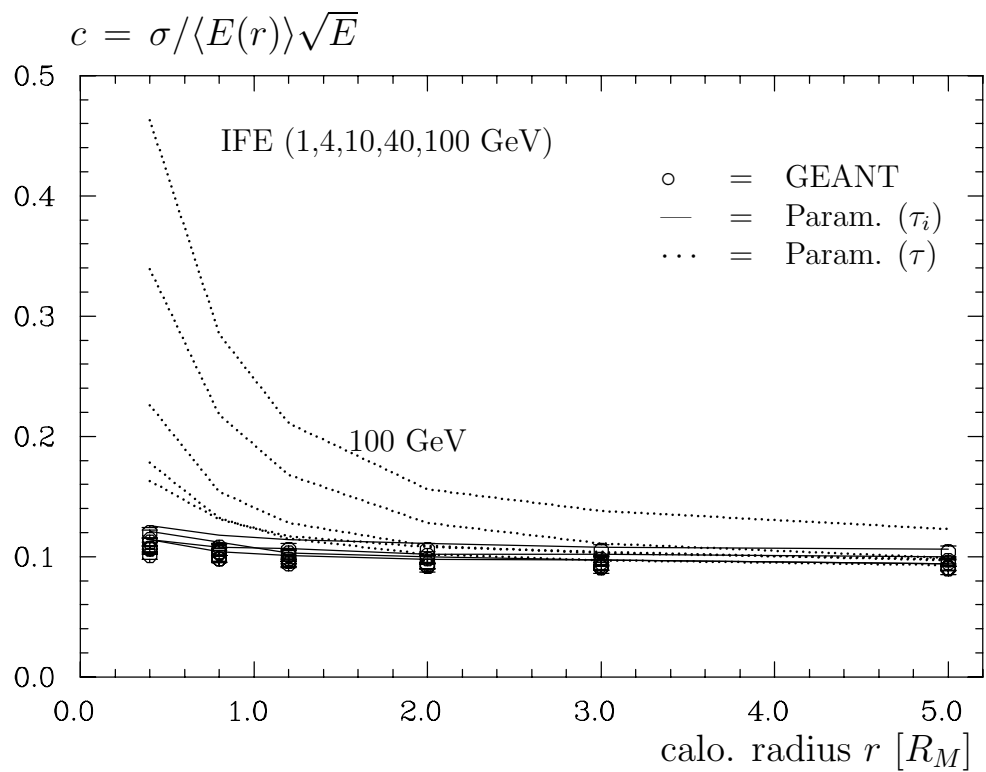
  \begin{figure}[p]\centering                                                 
   %
                                                                
   \caption[                                                                   
       Energy resolution vs. calorimeter radius in H1-IFE:
       --- ($\cdots$) with (without) long./rad. correlations
       (GEANT and param. simulation)]{}
     \label{clrsif}                                                             
  \end{figure}

 GEANT  and parameterized simulations of  mean radial profiles
 and their relative fluctuations,                                               
\begin{equation}                                                                
  \hat{\sigma}(t,r) \, = \, \frac{ \sigma_{RMS} }                               
  { \left\langle \frac{1}{dE(t)} \frac{dE(t,r)}{dr} \right\rangle },            
\label{sigmahat}                                                                
\end{equation}                                                                  
 are compared in Fig.\ref{cravif} and Fig.\ref{crinif}  
 for the H1 liquid argon calorimeter (IFE) for various energies.
 The influence of radial leakage on containment and energy                      
 resolution is demonstrated in Fig.\ref{clrcif} and                          
 Fig.\ref{clrsif}. The energy independence of the energy                            
 contained in a cylinder of radius $r$ is well reproduced by                  
 the parameterized simulations. The energy resolution as                        
 defined in Fig.\ref{clrsif} does not depend on radial leakage.
 As can be seen, this is correctly predicted by the parameterized
 simulation, when the correlation between the longitudinal                      
 and radial shower development is taken into account                            
 (by using $\tau_i$).

\section{Comparison with data}
                                                  
 We have compared parameterized simulations with test beam data                 
 from the H1 calorimeter, which is made of lead and liquid argon in the 
 electromagnetic sections \cite{h1prop,diss}. Modules of the inner forward 
 (IFE), the forward barrel (FB1), and the central barrel (CB2/CB3) calorimeters 
 have been studied. Electron beams in the energy range between 5 and 80 GeV 
 entered the stacks under angles of $11^{\circ}$ in the IFE and CB3,                        
 and under $35^{\circ}$ in the FB1 calorimeter in a test set-up at CERN.

 The energy resolution of the data can be described by                          
 \begin{equation}                                                               
   \frac{ \sigma (E)}{ \langle E \rangle }\, = \,                               
    \sqrt{                                                                      
           \frac{c^2}{\langle E \rangle }                                       
         + \frac{b^2}{\langle E \rangle^2 }                                     
         + \left( \frac{\sigma (p)}{ p }\right)^2  } .                          
   \label{gcresolu}                                                             
 \end{equation}                                                                 
 Here $c$ refers to the sampling fluctuations, $b$ considers                   
 the noise, and $\sigma (p) / p$ denotes the momentum                           
 resolution of the beam. In the Monte Carlo the momentum                        
 resolution was simulated explicitly. The electronic                       
 noise was taken into account by adding random trigger                          
 events to the simulated cell energies. The constant $c$, which is
 approximately $11\%$ for all modules, was used to                          
 simulate the sampling fluctuations.                                            

 The simulations were carried out with the H1 detector
 simulation program H1FAST \cite{kuhlen,rud92}.
 The algorithms described so far are part of this program, which is used
 for the mass production of Monte Carlo events in the H1
 detector at the HERA collider at DESY. To keep the required high precision
 of the parameterization also in complicated detector regions
 (cracks for example), the following has to be considered.
 If a shower develops partly inside cracks between                      
 adjacent modules, which in general cannot be approximated                      
 by a single effective medium, parameterizations will in                        
 general fail to reproduce measured signals.                                    
 In H1FAST\footnote{A stand alone version (called GFLASH 1.4) 
 running with GEANT and covering the same functionality is available for 
 distribution. Please contact one of the authors.} 
 we therefore do not parameterize showers, if they cross such boundaries.                        
 Only electromagnetic showers and sub-showers from hadronic
 interactions are parameterized which fit into one single stack.
                                                                                
 During analysis, a $3 \sigma$ noise cut was applied to both the 
 experimental data and H1FAST data
 at the cell level, and energy clusters were built from cells 
 containing energies above threshold. Energy distributions of the clusters 
 with maximum channel numbers are compared in Fig.\ref{ceclust} for all
 three modules considered. In addition, the energy in all other cells, not 
 belonging to the selected clusters, are also shown.
                                                                                
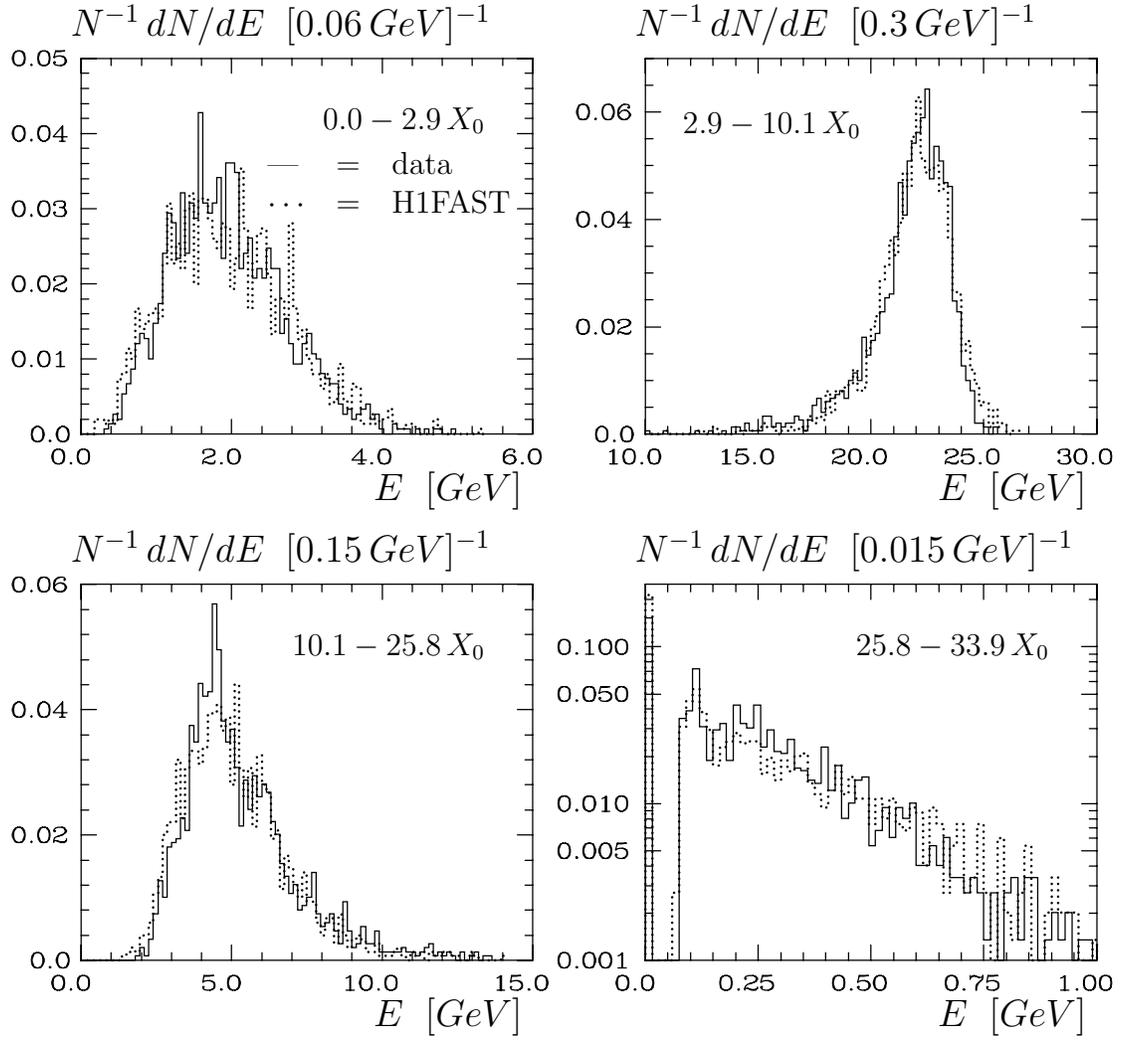
  \begin{figure}[p]\centering                                                 
   %
                                                                
    \caption[                                                                   
      Energy distributions in 4 subsequent layers of                         
      H1-CB3 for 30 GeV (data and H1FAST)]{}                                             
     \label{nimhist2}                                                           
  \end{figure}

 Longitudinal profiles are shown in Fig.\ref{clongife}                       
 for various energies in the IFE calorimeter. The mean                          
 profiles as well as the fluctuations are nearly                              
 indistinguishable between data and H1FAST. Energy                      
 distributions in individual longitudinal layers of CB3 are
 compared in Fig.\ref{nimhist2}                                              
 for 30 GeV incident electron energy, showing that not only the                    
 means and fluctuations but also the shape of the distributions                  
 are predicted correctly by the parameterized simulations.                      
                                                                                
 Fig.\ref{clatif10} compares lateral profiles in different                   
 shower depths in the IFE calorimeter at one energy,                            
 and in Fig.\ref{clatfb} lateral profiles in the FB1 calorimeter,
 summed over all longitudinal sections, are shown for various energies.                 
 There is good agreement in the peak distributions.                             
 The tails of the profiles are dominated by electronic noise.                   
                                                                                
 As shown so far, parameterized simulations can predict                          
 measured calorimeter signals very precisely, if the                             
 shower development is confined within one single calorimeter                   
 stack. Using the concept of partial parameterization as 
 described above, the                           
 influence of cracks on the measured signal can be reproduced                   
 as shown in Fig.\ref{cscan}. We have used test beam data
 scanning the crack between CB2 and CB3, which consist of
 two electromagnetic (CB2E, CB3E) and two hadronic stacks
 (CB2H, CB3H). The width of the crack is approximately 1 cm.
 Shown are the energies in the electromagnetic
 modules ($E_{CB2E}$, $E_{CB3E}$), the sum               
  of both ($E_{CBE} = E_{CB2E} + E_{CB3E}$), and the sum
 measured in the electromagnetic and hadronic modules
 ($E_{CB} = E_{CBE} + E_{CBH}$) as a function           
 of the beam impact position. All energies are normalized to            
 $E_{+20} \equiv E_{CB}(x_{calo} = 20 cm$).
 The energy lost while scanning the crack with                    
 a 30 GeV test beam extends to about                                           
 $40\%$, if only the electromagnetic sections are considered,                   
 and is still around $20\%$ if the hadronic modules are added.
 The agreement between data and                        
 partial parameterization is quite satisfactory.                                
                                                                                
 Of the various comparisons which were made \cite{diss},
 only a limited number is presented here. Other properties, 
 which are relevant for physics analysis with the H1 detector,
 like  $e/\pi$ separation, were studied \cite{diss,rud92} and 
 confirm the applicability of the fast simulation.

\section{Timing}
                                                                
 The CPU time reduction depends on the complexity of the                        
 geometry description and the cut off parameters in the                          
 detailed simulation as well as on the type of simulated                        
 event. Fully parameterized simulations of electromagnetic                     
 showers in a simple (box) geometry are about                                 
 7000 times faster at 100 GeV (900 at 1 GeV) compared                           
 with GEANT simulations of a detailed geometry and                         
 with low energy cuts ($e$-cut$=200$ keV, $\gamma$-cut$=10$ keV).                        
                                                                                
 In the framework of the H1 simulation program, partial                          
 parameterization of electromagnetic showers is performed                       
 as described above, together with individual tracking of                        
 hadrons and termination of low energy particles
 (see also \cite{kuhlen,rud92}). The gain factors for
 30 GeV  showers in the H1 detector (including detailed 
 simulations  of tracker volumes) are 200 for electrons and 
 25 in  case of hadronic showers. 
 Medium energy cuts  ($e$-cut$=1$ MeV, $\gamma$-cut$=200$ keV) were 
 used in the corresponding detailed simulations.         
 Complete detector simulations  of HERA events 
 (ep scattering at $\sqrt{s}$ = 314 GeV)  require at least 
 10 times less CPU time using partial parameterization.
                                                                               
\section{Conclusions}
                                                           
 We have developed parameterizations of electromagnetic                         
 showers for different materials and sampling geometries.                       
 Shower to shower fluctuations                         
 and correlations are taken into account consistently,                          
 as well as correlations between the longitudinal and                               
 radial shower development. Comparisons with data have                          
 shown that parameterized simulations are able to predict                       
 measured calorimeter signals with an acceptable precision.
 Using the methods described above, the energy resolution is
 reproduced at the level of $\pm 0.5 \%$. The energy 
 deposited in longitudinal and lateral layers is predicted
 with a precision of typically $\pm 1.5\%$ for both, the
 means and the fluctuations. Using partial parameterizations,
 the energy measured in electromagnetic (and hadronic) modules 
 differs by an amount of $1.7\%$ ($9\%$), if the beam enters
 directly into a crack. The parameterizations presented here provide 
 a fast and precise algorithms for large scale Monte Carlo 
 production of events for physics analysis.                              
%
%
%

\vspace*{10mm}
\appendix                                                                       
\section{Summary of formulae}                                                   
\label{appendix}                                                                
 \subsection{Homogeneous Media}                                                  
 \label{a_hom}                                                                  
  \subsubsection{Average longitudinal profiles}                                 
  \label{a_hom_long_ave}                                                        
  \begin{eqnarray*}                                                             
   T_{hom} & = & \ln y - 0.858 \\                                               
   \alpha_{hom}    & = & 0.21+ (0.492 + 2.38/Z) \ln y                           
  \end{eqnarray*}                                                               
%
  \subsubsection{Fluctuated longitudinal profiles}                              
  \label{a_hom_long_ind}                                                        
  \begin{eqnarray*}                                                             
   \langle \ln T_{hom} \rangle & = &\ln( \ln y -0.812 ) \\                      
   \sigma(\ln T_{hom}) & = & ( -1.4 + 1.26 \ln y )^{-1}\\                       
   \langle \ln \alpha_{hom}\rangle & = &                                        
       \ln \left( 0.81 + (0.458 + 2.26/Z) \ln y \,\right) \\                    
   \sigma(\ln \alpha_{hom}) & = & ( -0.58+ 0.86 \ln y )^{-1}\\                  
   \rho(\ln T_{hom}, \ln \alpha_{hom}) & = & 0.705 -0.023 \ln y                 
  \end{eqnarray*}                                                               
%
  \subsubsection{Average radial profiles}                                       
  \label{a_hom_rad_ave}                                                         
  \begin{eqnarray*}                                                             
     R_{C,hom}(\tau) & = &                                                      
            z_1 + z_2 \tau \\                                                   
     R_{T,hom}(\tau) & = &                                                      
        k_1 \{ \exp (k_3(\tau -k_2)) + \exp (k_4(\tau -k_2)) \} \\              
     p_{hom}(\tau) & = &                                                        
        p_1  \exp \left\{ \frac{p_2-\tau}{p_3} -                                
        \exp \left(   \frac{p_2-\tau}{p_3} \right) \right\}\\                   
          & \mbox{with} & \\                                                    
     z_1 & = & 0.0251 + 0.00319 \ln E       \\                                  
     z_2 & = & 0.1162 + -0.000381 Z             \\                              
     k_1 & = & 0.659  + -0.00309 Z             \\                               
     k_2 & = & 0.645                         \\                                 
     k_3 & = & -2.59                         \\                                 
     k_4 & = & 0.3585 + 0.0421 \ln E         \\                                 
     p_1 & = & 2.632  + -0.00094 Z             \\                               
     p_2 & = & 0.401  + 0.00187 Z             \\                                
     p_3 & = & 1.313  + -0.0686 \ln E                                           
  \end{eqnarray*}                                                               
%
  \subsubsection{Fluctuated radial profiles}                                    
  \label{a_hom_rad_ind}                                                         
  \begin{eqnarray*}                                                             
     \tau_i & = & \frac{t}{\langle t \rangle_i}                                 
         \frac{\exp(\langle \ln \alpha \rangle )}                               
         {\exp(\langle \ln \alpha \rangle )-1} \\                               
     N_{Spot} & = & 93 \ln(Z) E^{0.876}\\                                       
     T_{Spot} & = & T_{hom} ( 0.698 + 0.00212 Z ) \\                            
     \alpha_{Spot} & = & \alpha_{hom} ( 0.639 + 0.00334 Z )                     
  \end{eqnarray*}                                                               
                                                                                
 \subsection{Sampling Calorimeters}                                             
 \label{a_sam}                                                                  
%
  \subsubsection{Material and geometry parameters}                              
  \label{a_sam_mat}                                                             
  \begin{eqnarray*}                                                             
   w_i           & = &                                                          
     \frac{\rho_i d_i}{\sum_{j} \rho_j d_j}                                     
     \,\, \mbox{ $(\rho =$~density$)$} \\                                         
   Z_{eff}           & = &                                                      
     \sum_{i} w_i Z_i \; \; \\                                                  
   A_{eff}           & = &                                                      
     \sum_{i} w_i A_i \\                                                        
   \frac{1}{X_{0,eff}}           & = &                                          
     \sum_{i} \frac{w_i}{X_{0,i}} \\                                            
     \frac{1}{R_{M,eff}}           & = &                                        
     \frac{1}{E_s} \sum_{i} \frac{w_i E_{c,i}    }{X_{0,i}}
     \,\, \mbox{ $(E_s = 21.2$~MeV$)$} \\                   
   E_{c,eff}                & = &                                               
     X_{0,eff}  \sum_{i} \frac{w_i E_{c,i}    }{X_{0,i}}\\                      
    F_S & = & \frac{X_{0,eff}}{d_a + d_p} \\                                    
    \hat{e} & = & \frac{1}{1+0.007(Z_p-Z_a)}                                    
  \end{eqnarray*}                                                               
%
  \subsubsection{Average longitudinal profiles}                                 
  \label{a_sam_long_ave}                                                        
  \begin{eqnarray*}                                                             
    T_{sam} & = & T_{hom} -0.59 F_S^{-1} -0.53 (1-\hat{e}) \\                   
    \alpha_{sam} & = & \alpha_{hom} -0.444 F_S^{-1} \\                          
  \end{eqnarray*}                                                               
%
  \subsubsection{Fluctuated longitudinal profiles}                              
  \label{a_sam_long_ind}                                                        
  \begin{eqnarray*}                                                             
   \langle \ln T_{sam} \rangle & = &                                            
      \ln \left(  \exp(\langle \ln T_{hom} \rangle )                            
      -0.55 F_S^{-1} -0.69 (1 - \hat{e}) \right) \\                             
   \sigma(\ln T_{sam}) & = & ( -2.5 + 1.25 \ln y )^{-1}\\                       
   \langle \ln \alpha_{sam} \rangle & = &                                       
      \ln \left( \exp( \langle \ln \alpha_{hom} \rangle )                       
      -0.476 F_S^{-1} \right) \\                                                
   \sigma(\ln \alpha_{sam}) & = & ( -0.82+ 0.79 \ln y )^{-1}\\                  
   \rho(\ln T_{sam}, \ln \alpha_{sam}) & = & 0.784 -0.023 \ln y                 
  \end{eqnarray*}                                                               
%
  \subsubsection{Average radial profiles}                                       
  \label{a_sam_rad_ave}                                                         
  \begin{eqnarray*}                                                             
    R_{C,sam} & = & R_{C,hom}                                                   
      -0.0203  ( 1-\hat{e} )                                                    
      +0.0397  F_S^{-1} \exp( -\tau) \\                                         
    R_{T,sam} & = & R_{T,hom}                                                   
      -0.14 ( 1-\hat{e} )                                                       
      -0.495 F_S^{-1} \exp( -\tau) \\                                           
    p_{sam} & = & p_{hom} +                                                     
      ( 1-\hat{e} )                                                             
      ( 0.348 -0.642 F_S^{-1} \exp(-(\tau-1)^2)  )                              
  \end{eqnarray*}                                                               
%
  \subsubsection{Fluctuated radial profiles}                                    
  \label{a_sam_rad_ind}                                                         
  \begin{eqnarray*}                                                             
     \tau_i & = & \frac{t}{\langle t \rangle_i}                                 
         \frac{\exp(\langle \ln \alpha \rangle )}                               
         {\exp(\langle \ln \alpha \rangle )-1} \\                               
     N_{Spot} & = & \frac{10.3}{c} E^{0.959}
            \;\;\;\; (\frac{\sigma}{E}=\frac{c}{\sqrt{E}}) \\   
     T_{Spot} & = & T_{hom} ( 0.813 + 0.0019  Z ) \\                            
     \alpha_{Spot} & = & \alpha_{hom} ( 0.844 + 0.0026  Z )                     
  \end{eqnarray*}                                                               
\clearpage                                                                      
\listoffigures
  \begin{figure}[p]\centering                                                 
                                                                
    \caption[                                                                   
        Depths of the shower maxima as a function of $y$ (GEANT).]{}  
     \label{p_thom}                                                             
  \end{figure}                                                                  
\clearpage
  \begin{figure}[p]\centering                                                 
   %
                                                                
    \caption[Shape parameter $\alpha$ as a function of $y$ (GEANT).]{}
     \label{p_ahom}                                                             
  \end{figure}                                                                  
  \begin{figure}[p]\centering                                                 
   %
                                                                
    \caption[                                                                   
      Parameters of longitudinal profiles                    
      in homogeneous media (GEANT)]{}                                          
     \label{fitlhom}                                                            
  \end{figure}                                                                  
  \begin{figure}[p]\centering                                                 
   %
                                                                
    \caption[                                                                   
         Mean longitudinal profiles and their                        
         fluctuations in a lead glass calorimeter (GEANT and param. simulation)]{}                  
     \label{longsf5}                                                            
  \end{figure}                                                                  
  \begin{figure}[p]\centering                                                 
   %
                                                                
    \caption[                                                                   
       Parameters of individual longitudinal energy profiles
       in sampling calorimeters (GEANT)]{}                                     
     \label{fitlsam}                                                            
   \end{figure}                                                                 
  \begin{figure}[p]\centering                                                 
   %
                                                                
    \caption[                                                                   
         Mean longitudinal profiles and their                       
         fluctuations in the H1 lead liquid argon                  
         calorimeter IFE (H1-IFE) (GEANT and param. simulation)]{}
     \label{clinif}                                                             
  \end{figure}                                                                  
  \begin{figure}[p]\centering                                                 
   %
                                                                
    \caption[                                                                   
       Energy containment vs. calorimeter depth
       in H1-IFE (GEANT and param. simulation)]{}                                
     \label{cllcif}                                                             
  \end{figure}                                                                  
\clearpage
  \begin{figure}[p]\centering                                                 
   %
                                                                
    \caption[                                                                   
      Energy resolution vs. calorimeter depth in H1-IFE (GEANT and 
      param. simulation)]{}
     \label{cllsif}                                                             
  \end{figure}                                                                  
\clearpage
  \begin{figure}[p]\centering                                                 
   %
                                                                
    \caption[                                                                   
      Mean radial profiles in uranium at various calorimeter                    
         depth (GEANT). The two component Ansatz is indicated                  
         for the depth $1-2 X_0$]{}
     \label{hradx1}                                                             
  \end{figure}                                                                  
\clearpage
  \begin{figure}[p]\centering                                                 
   %
                                                                
    \caption[                                                                   
       Shower depth and material dependence of radial shower                    
       parameters]{}
     \label{hrzrkp}                                                             
  \end{figure}                                                                  
\clearpage
  \begin{figure}[p]\centering                                                 
    %
                                                               
    \caption[                                                                   
        Tungsten, 40 GeV.                                                       
         Left figure: Decomposition of the two sources of radial                
         shape fluctuations. Right figure: Comparison of energy                 
         and spot density distributions]{}                                       
     \label{sigbyew}                                                            
  \end{figure}                                                                  
  \begin{figure}[p]\centering                                                 
                                                                
    \caption[                                                                   
         Mean radial profiles and their fluctuations                
         in lead (40 GeV) and uranium (100 GeV) (GEANT and param. simulation)]{}
     \label{crinhom3}                                                           
  \end{figure}                                                                  
%
%
  \begin{figure}[p]\centering                                                 
   %
                                                                
    \caption[                                                                   
         Top: Mean integrated radial profiles in lead glass.
         Bottom: Corresponding shape fluctuations:
         --- ($\cdots$) with (without) long./rad. correlations (GEANT and 
         param. simulation)]{}
     \label{radsf5}                                                             
  \end{figure}                                                                  
\clearpage
  \begin{figure}[p]\centering                                                 
   %
                                                                
    \caption[                                                                   
       Total number of spots required to reproduce radial shape fluctuations in
       sampling calorimeters]{}                                
     \label{nspots}                                                             
  \end{figure}                                                                  
  \begin{figure}[p]\centering                                                 
   %
                                                                
    \caption[                                                                   
         Mean radial profiles in H1-IFE                  
         (GEANT and param. simulation)]{}
     \label{cravif}                                                             
  \end{figure}                                                                  
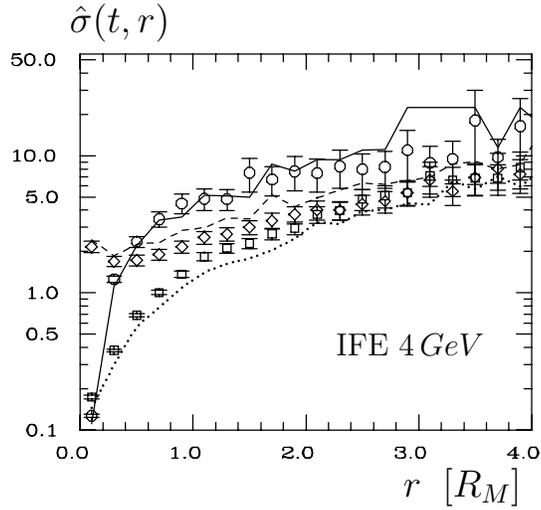
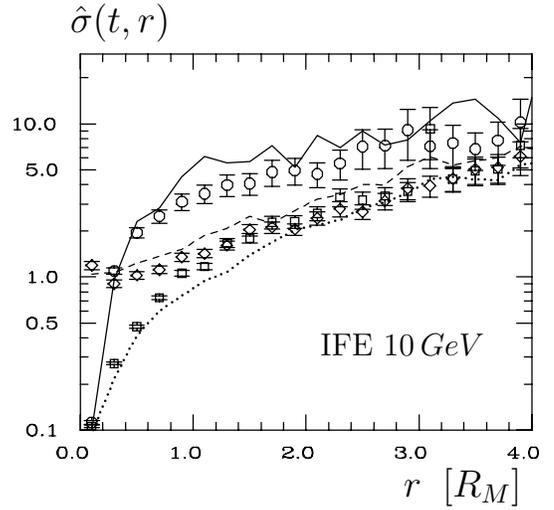
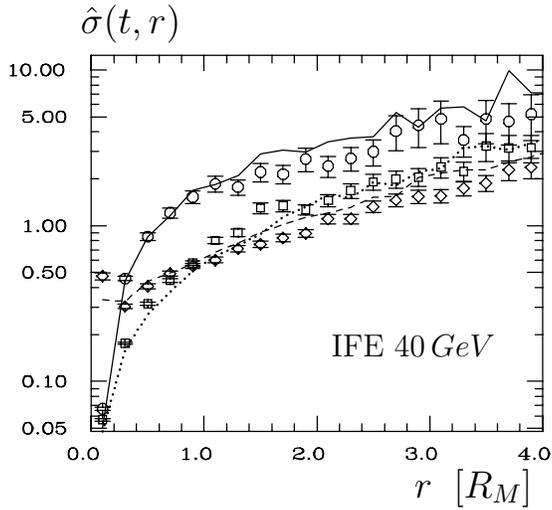
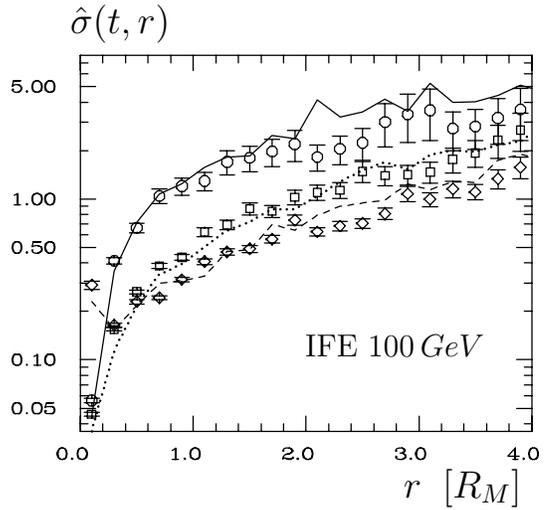
  \begin{figure}[p]\centering                                                 
                                                                
    \caption[                                                                   
         Radial shape fluctuations in H1-IFE            
         (GEANT and param. simulation)]{}
     \label{crinif}                                                             
  \end{figure}                                                                  
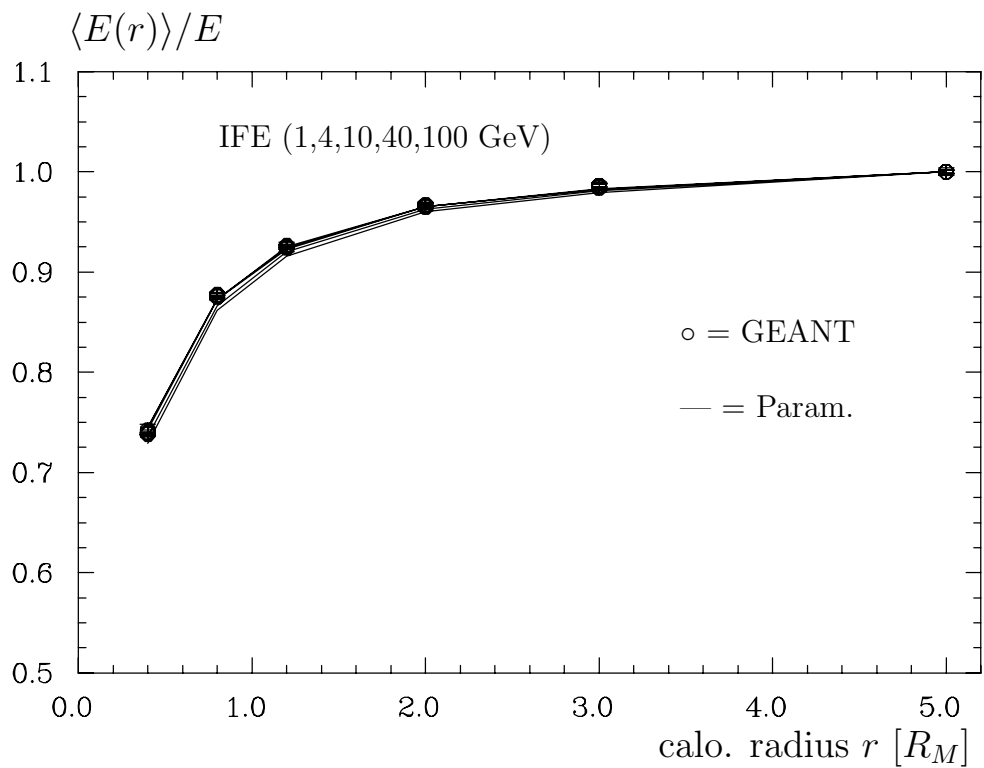
  \begin{figure}[p]\centering                                                 
   %
                                                                
    \caption[                                                                   
      Energy containment vs. calorimeter radius in H1-IFE
      (GEANT and param. simulation)]{}
     \label{clrcif}                                                             
  \end{figure}                                                                  
\clearpage
\clearpage                                                                      
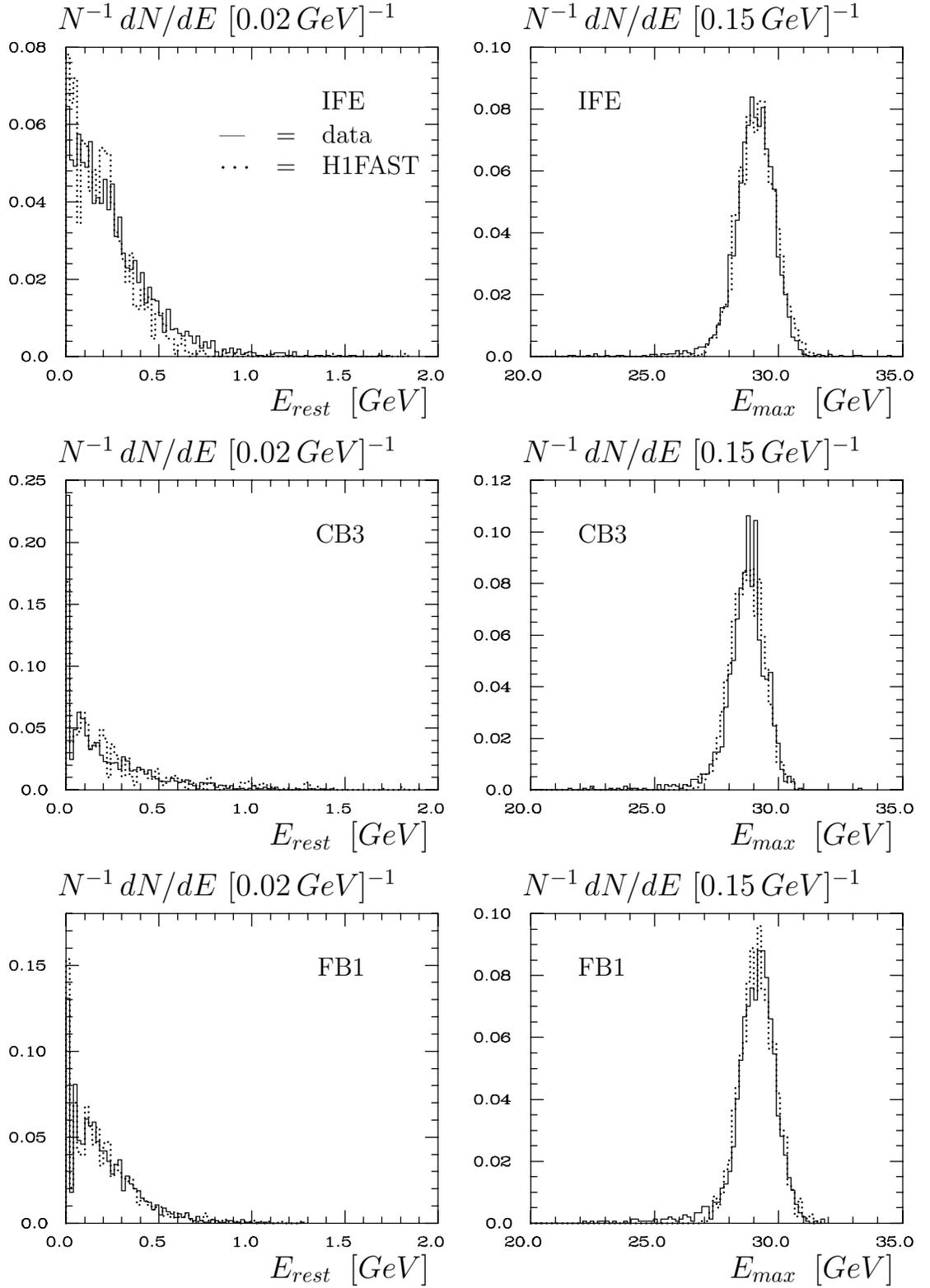
  \begin{figure}[p]\centering                                                 
   %
                                                                
    \caption[                                                                   
       Energy distributions of the cluster with the maximum number of channels                 
         $(E_{max})$ and of all other channels $(E_{rest})$ for
         30 GeV showers in three different modules, IFE, CB3, and FB1 of the H1                     
         lead liquid argon calorimeter (data and H1FAST)]{}
     \label{ceclust}                                                            
  \end{figure}                                                                  
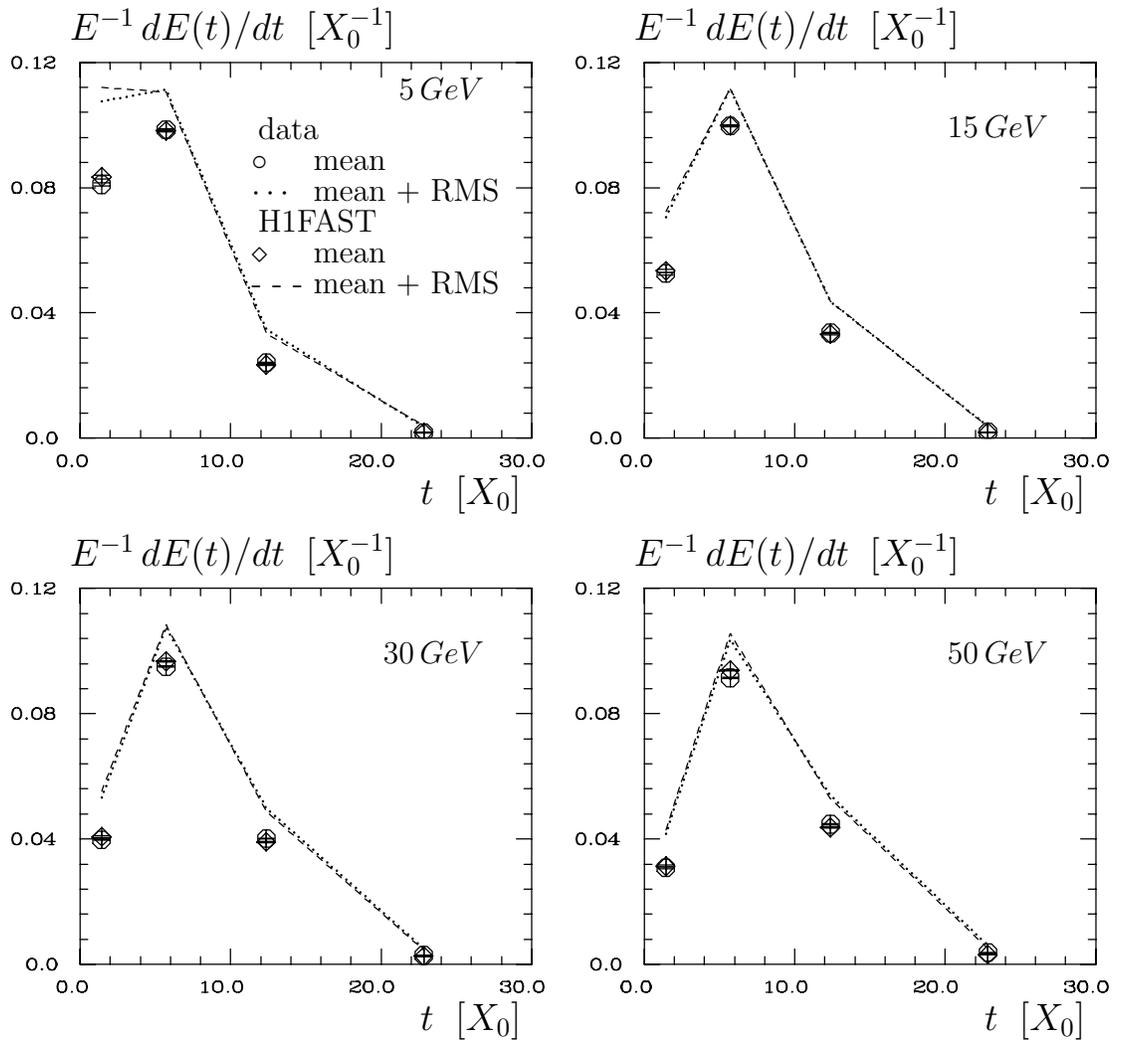
  \begin{figure}[p]\centering                                                 
   %
                                                                
    \caption[                                                                   
       Mean longitudinal profiles and fluctuations in                           
       H1-IFE (data and H1FAST)]{}
     \label{clongife}                                                           
  \end{figure}                                                                  
\clearpage
\clearpage
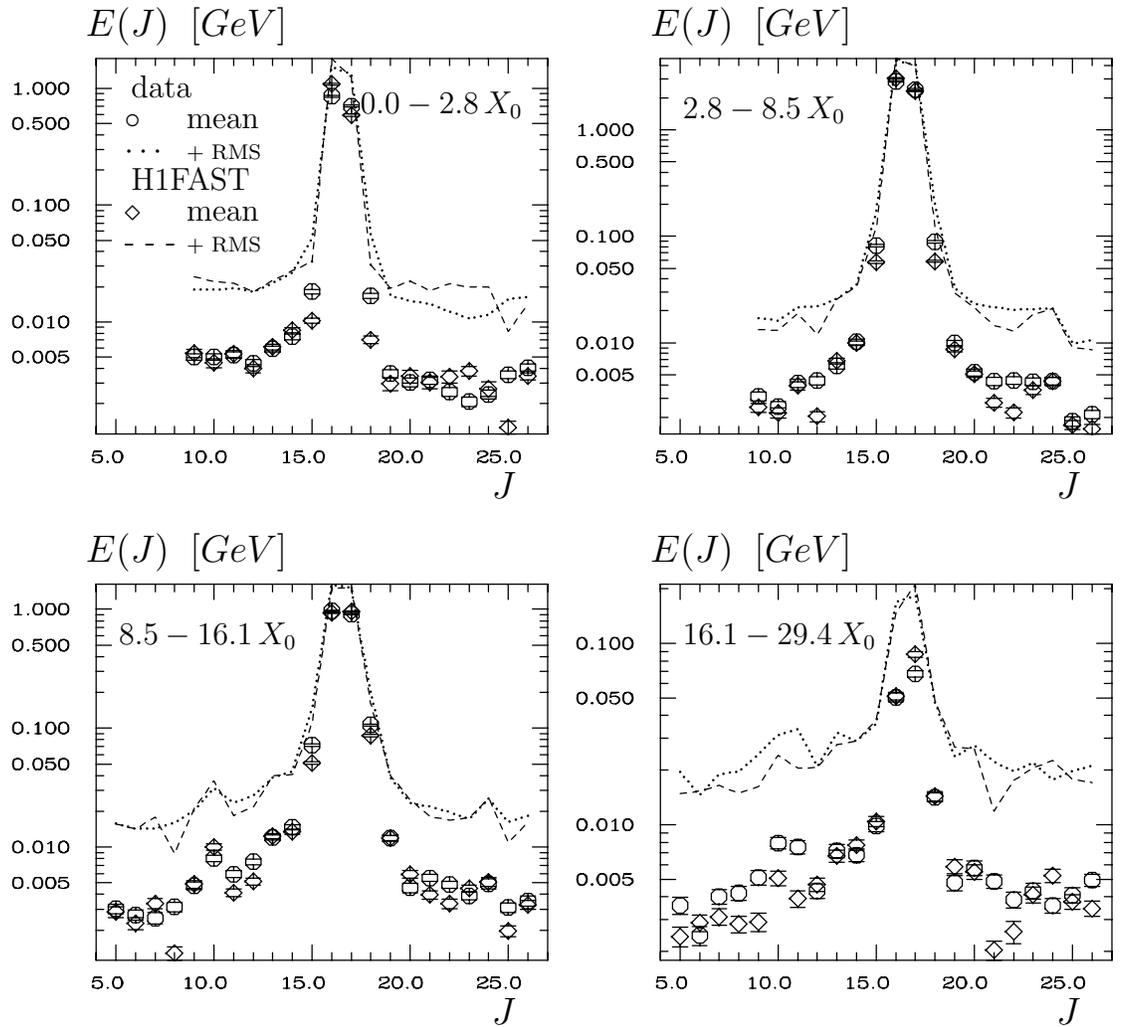
  \begin{figure}[p]\centering                                                 
   %
                                                                
    \caption[                                                                   
      Mean lateral profiles and fluctuations                                    
      at different depths in H1-IFE for 10 GeV (data and H1FAST).            
      The coordinates $J$ correspond to transverse pad numbers]{}
     \label{clatif10}                                                           
  \end{figure}                                                                  
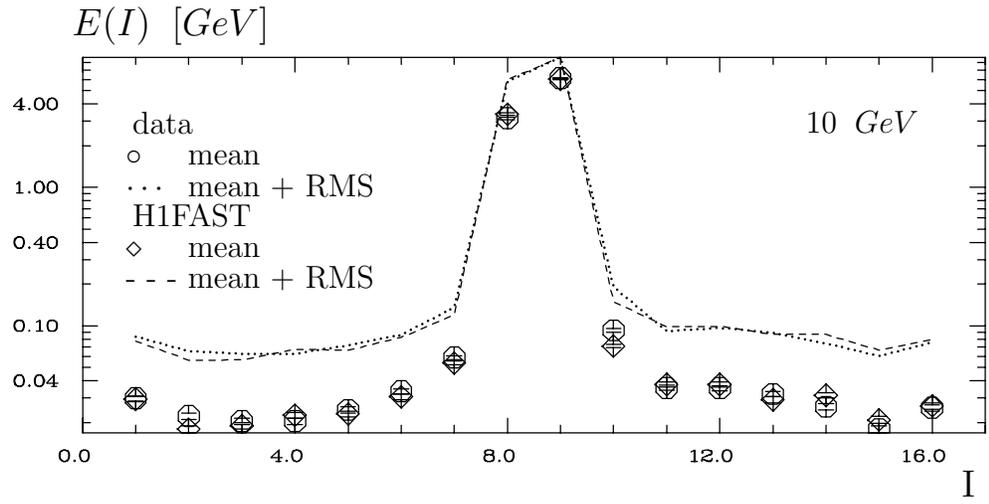
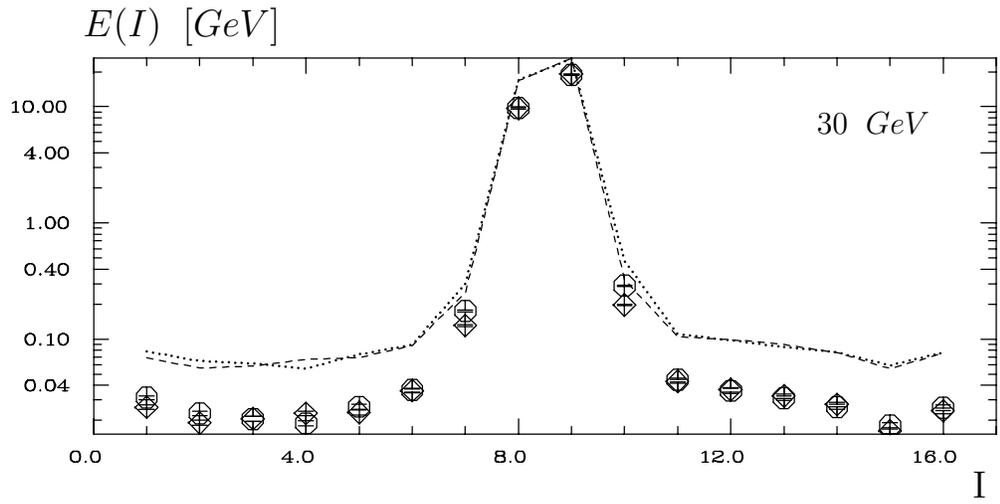
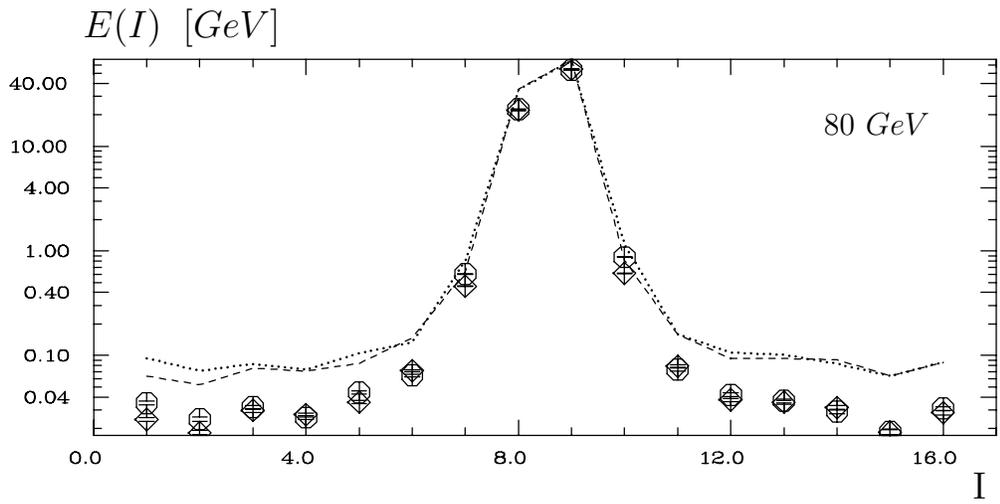
  \begin{figure}[p]\centering                                                 
   %
                                                                
    \caption[                                                                   
       Mean lateral profiles and fluctuations in H1-FB1 (data and H1FAST). 
       The coordinates $I$ correspond to transverse pad numbers]{}
    \label{clatfb}                                                              
  \end{figure}                                                                  
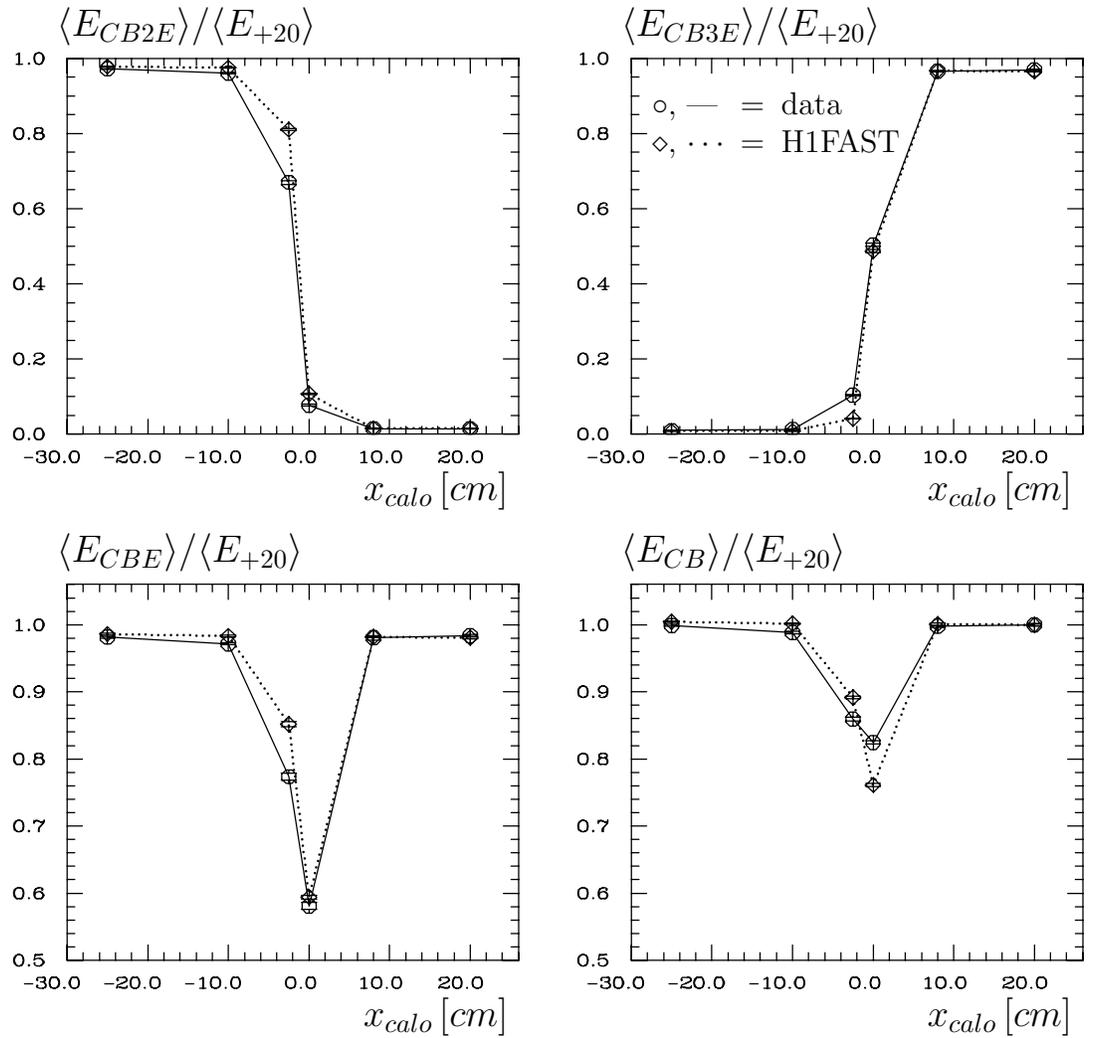
  \begin{figure}[p]\centering                                                 
   %
                                                                
    \caption[
       Normalized energies as measured across a crack between
       the modules H1-CB2 and H1-CB3 (data and H1FAST)]{}
     \label{cscan}                                                              
  \end{figure}                                                                  
\clearpage                                                                      
\pagestyle{empty}                                                               

\newpage                                                                               
\begin{thebibliography}{20}                                                     
 \newcounter{bibcounter}                                                        
 \setcounter{bibcounter}{0}                                                     
%
   \addtocounter{bibcounter}{1}                                                 
   \bibitem[\thebibcounter]{longo} E.~Longo and I.~Sestili,                     
       Nucl.~Instrum.~Meth.~128, 283, (1975).                                  
%
   \addtocounter{bibcounter}{1}                                                 
   \bibitem[\thebibcounter]{bock} R.~K.~Bock et al.,                            
       Nucl.~Instrum.~Meth.~186, 533, (1981); \\                              
       M.~della Negra, Scripta Phys.~23, 469, (1981).                           
%
   \addtocounter{bibcounter}{1}                                                 
   \bibitem[\thebibcounter]{hayashide} Y.~Hayashide et al.,                     
       CDF Note 287, Batavia, IL (1985).                                        
%
   \addtocounter{bibcounter}{1}                                                 
   \bibitem[\thebibcounter]{badier} J.~Badier and M.~Bardadin-%
       Otwinowska, ALEPH 87--9, EMCAL 87--1, Geneva (1987).                     
%
   \addtocounter{bibcounter}{1}                                                 
   \bibitem[\thebibcounter]{nim90} G.~Grindhammer, M.~Rudowicz, and              
       S.~Peters, Nucl.~Instrum.~Meth.~A290, 469, (1990).
%
   \addtocounter{bibcounter}{1}                                                 
   \bibitem[\thebibcounter]{diss} S.~Peters,                                    
       PhD thesis, University of Hamburg;                                 
       MPI--PhE/92--13 (1992).
%
   \addtocounter{bibcounter}{1}                                                 
   \bibitem[\thebibcounter]{geant} R.~Brun et al.,                              
       GEANT3 User's Guide. CERN--DD/EE 84--1, Geneva (1986).                   
%
   \addtocounter{bibcounter}{1}                                                 
   \bibitem[\thebibcounter]{flauger} W.~Flauger,                                
               Nucl.~Instrum.~Meth.~A241, 72, (1985).                          
%
   \addtocounter{bibcounter}{1}                                                 
   \bibitem[\thebibcounter]{wigmans} R.~Wigmans,
               Nucl.~Instrum.~Meth.~A259, 389, (1987).                 
%
   \addtocounter{bibcounter}{1}                                                 
   \bibitem[\thebibcounter]{rossi} B.~Rossi,                                    
               Prentice Hall, New York, (1952).                                  
%
   \addtocounter{bibcounter}{1}                                                 
   \bibitem[\thebibcounter]{dov} O.~I.~Dovzhenkko and                           
               A.~A.~Pommanskii,                                                
               Sov.~Phys.~JETP Vol.~18, Numb.~1, (1964).                     
%
   \addtocounter{bibcounter}{1}                                                 
   \bibitem[\thebibcounter]{delpeso} J.~del Peso and E.~Ros,
       Nucl.~Instrum.~Meth.~A295, 330, (1990).
%
   \addtocounter{bibcounter}{1}                                                 
   \bibitem[\thebibcounter]{h1prop} H1 Collab.,                           
       Technical Proposal for the H1 Detector, DESY, (1986); \\
       H1 Collab., I.~Abt et al., Nucl.~Instrum.~Meth.~A386, 310 and A386, 348, (1997).  
%
   \addtocounter{bibcounter}{1}                                                 
   \bibitem[\thebibcounter]{ako77} G.~A.~Akopdjanov et al.,
       Nucl.~Instrum.~Meth.~140, 441, (1977).
%
   \addtocounter{bibcounter}{1}                                                 
   \bibitem[\thebibcounter]{abs79} G.~Abshire et al.,
       Nucl.~Instrum.~Meth.~164, 67, (1979).
%
   \addtocounter{bibcounter}{1}                                                 
   \bibitem[\thebibcounter]{fer88} G.~Ferri et al.,
       Nucl.~Instrum.~Meth.~A273, 123, (1988).
%
   \addtocounter{bibcounter}{1}                                                 
   \bibitem[\thebibcounter]{peso89} J.~del~Peso and E.~Ros,
       Nucl.~Instrum.~Meth.~A276, 456, (1989).
%
   \addtocounter{bibcounter}{1}                                                 
   \bibitem[\thebibcounter]{sicapo} SICAPO Collab., E.~Borchi et al.,
       Nucl.~Phys.~Proc.~Suppl.~23A, 119, (1991).
%
   \addtocounter{bibcounter}{1}                                                 
   \bibitem[\thebibcounter]{deangelis} A.~de Angelis and P.~A.~Palazzi,
       Nucl.~Instrum.~Meth.~A271, 455, (1988).
%
%
   \addtocounter{bibcounter}{1}                                                 
   \bibitem[\thebibcounter]{kuhlen} M.~Kuhlen,                                  
       in Proceedings of the XXVI International                
       Conference on High Energy Physics, Dallas (1992).
%
   \addtocounter{bibcounter}{1}                                                 
   \bibitem[\thebibcounter]{rud92} M.~Rudowicz,
       PhD thesis, University of Hamburg;                                 
       MPI--PhE/92--14 (1992). 
%
\end{thebibliography}
\end{document}